\documentclass[table,amsmath,amssymb,aps,prl,superscriptaddress,twocolumn,floatfix]{revtex4-1}

\usepackage{graphicx,
    hyperref,
    microtype,
    siunitx,
    tabu,
    multirow,
    textcomp,
    tikz,
    xspace
}

\def\unit #1 #2 {\SI{#1}{#2}\xspace}
\def\mum{\micro m\xspace}

\def\ele#1#2{\textsuperscript{#1}#2}

\DeclareSIUnit\gauss{G}
\usepackage{blindtext}
\hypersetup{colorlinks=true, linkcolor=blue,citecolor=blue,urlcolor=blue}

\newcommand{\myref}[2][]{\hyperref[#2]{Fig.~\ref*{#2}#1}}
\newcommand{\Myref}[2][]{\hyperref[#2]{Figure~\ref*{#2}#1}}
\newcommand{\Mytabref}[2][]{\hyperref[#2]{Table~\ref*{#2}#1}}

\begin{document}

\preprint{double_BEC}
\author{A.~Trautmann}
\thanks{These authors contributed equally to this work.}
\affiliation{%
 Institut f\"{u}r Quantenoptik und Quanteninformation, \"Osterreichische Akademie der Wissenschaften, 6020 Innsbruck, Austria
}%
\author{P. Ilzh\"ofer}
\thanks{These authors contributed equally to this work.}
\affiliation{%
 Institut f\"{u}r Quantenoptik und Quanteninformation, \"Osterreichische Akademie der Wissenschaften, 6020 Innsbruck, Austria
}%
\affiliation{%
 Institut f\"{u}r Experimentalphysik und Zentrum f\"{u}r Quantenoptik,\\ Universit\"{a}t Innsbruck, Technikerstra\ss e 25, 6020 Innsbruck, Austria
}%
\author{G.~Durastante}
\affiliation{%
 Institut f\"{u}r Quantenoptik und Quanteninformation, \"Osterreichische Akademie der Wissenschaften, 6020 Innsbruck, Austria
}%
\affiliation{%
 Institut f\"{u}r Experimentalphysik und Zentrum f\"{u}r Quantenoptik,\\ Universit\"{a}t Innsbruck, Technikerstra\ss e 25, 6020 Innsbruck, Austria
}%
\author{C.~Politi}
\affiliation{%
 Institut f\"{u}r Quantenoptik und Quanteninformation, \"Osterreichische Akademie der Wissenschaften, 6020 Innsbruck, Austria
}%
\author{M.~Sohmen}
\affiliation{%
 Institut f\"{u}r Quantenoptik und Quanteninformation, \"Osterreichische Akademie der Wissenschaften, 6020 Innsbruck, Austria
}%
\affiliation{%
 Institut f\"{u}r Experimentalphysik und Zentrum f\"{u}r Quantenoptik,\\ Universit\"{a}t Innsbruck, Technikerstra\ss e 25, 6020 Innsbruck, Austria
}%
\author{M.~J.~Mark}
\affiliation{%
 Institut f\"{u}r Quantenoptik und Quanteninformation, \"Osterreichische Akademie der Wissenschaften, 6020 Innsbruck, Austria
}%
\affiliation{%
 Institut f\"{u}r Experimentalphysik und Zentrum f\"{u}r Quantenoptik,\\ Universit\"{a}t Innsbruck, Technikerstra\ss e 25, 6020 Innsbruck, Austria
}%

\author{F. Ferlaino}
\affiliation{%
 Institut f\"{u}r Quantenoptik und Quanteninformation, \"Osterreichische Akademie der Wissenschaften, 6020 Innsbruck, Austria
}%
\affiliation{%
 Institut f\"{u}r Experimentalphysik und Zentrum f\"{u}r Quantenoptik,\\ Universit\"{a}t Innsbruck, Technikerstra\ss e 25, 6020 Innsbruck, Austria
}%

\title{
Dipolar Quantum Mixtures of Erbium and Dysprosium Atoms
}
\date{June 2018}

\begin{abstract}
We report on the first realization of heteronuclear dipolar quantum mixtures of  highly magnetic erbium and dysprosium atoms. With a versatile experimental setup, we demonstrate binary Bose-Einstein condensation in five different Er-Dy isotope combinations, as well as one Er-Dy Bose-Fermi mixture. Finally, we present first studies of the interspecies interaction between the two species for one mixture.
\end{abstract}
\maketitle

In recent years, the field of atomic dipolar quantum gases has witnessed an impressive expansion as researchers have made substantial headway in using and controlling a novel class of atoms, the highly magnetic rare-earth species.  
Since the first experimental successes in creating Bose and Fermi quantum gases of Dy \cite{Lu2011,Lu2012} and Er \cite{Aikawa2012,Aikawa2014a}, fascinating dipolar phenomena and novel many-body quantum phases, including Fermi surface deformation \cite{Aikawa2014}, quantum-stabilized droplet states \cite{Kadau2016,Chomaz2016,Schmitt2016}, and roton quasi-particles \cite{Chomaz2018} have been observed. 
Remarkably, in Dy and Er, the intriguing  physics within reach comes with comparatively simple experimental approaches to achieve quantum degeneracy. Several research groups have either recently reported on new experimental realizations of quantum gases with Dy \cite{Maier2014,Lucioni2018} or Er \cite{Ulitzsch2017}, or are actively pursuing it \cite{Dreon2017,Ravensbergen2018}.

An alternative route to access dipolar physics is provided by ultracold gases of heteronuclear molecules \cite{Moses2016}, offering an electric dipole moment. Up to now, ultracold molecules have been created from non-dipolar binary quantum mixtures of alkali atoms \cite{Ni2008,Takekoshi2014, Molony2014}. Beside molecule creation, such quantum mixtures have been used as powerful resources to realize a broad class of many-body quantum states (e.\,g.~\cite{Ho1996,Ospelkaus2006,Will2011,Heinze2011,Spethmann2012,Jorgensen2016,Hu2016,Rentrop2016}), in which intra- and interspecies short-range contact interactions are at play.

In the experiment described in this Letter, we merge for the first time the physics of heteronuclear mixtures with the one of dipolar quantum gases. Our motivations to create quantum mixtures of two different dipolar species, Er  and Dy, are numerous.  
First, there is a clear fundamental interest in combining these two species to perform comparative studies and deeper elucidate the complex scattering and many-body physics with increasing strength of the magnetic and orbital anisotropic interaction. Second, switching from contact-interacting to dipolar mixtures, also the coupling between the two components acquires a unique anisotropic and long-range character, due to strong interspecies dipole-dipole interaction (DDI). The emergent physical richness of the system has only begun to be uncovered by theory. Recent studies include the prediction of anisotropic boundaries in the dipolar immiscibility-miscibility phase diagram \cite{Gligoric2010,Kumar2017}, roton immiscibility \cite{Wilson2012}, vortex lattice formation \cite{Kumar2017a}, and  impurity physics in dipolar quantum droplets \cite{Wenzel2017}.
Last, mixtures composed of two different magnetic species serve as an ideal platform to produce ground-state molecules with both an electric and magnetic dipole moment, offering novel degrees of control \cite{Abrahamsson2007,Rvachov2017}.

\begin{figure}[t]
    \centering
    \includegraphics[]{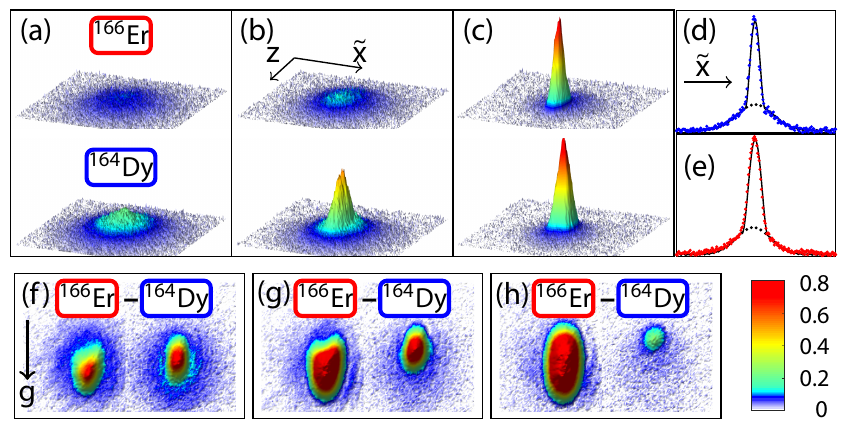}
    \caption{Binary Bose-Einstein condensation in a \ele{166}{Er}-\ele{164}{Dy} mixture. (a-c) Pairs of TOF absorption images at different evaporation stages, showing (a) a thermal mixture at about \unit 180 nK , (b) an Er cloud at the onset of condensation coexisting with a thermal Dy gas at about \unit 80 nK , and (c) the binary dipolar BEC with total atom numbers $N=\num{3.4e4}(\num{2.6e4})$ for Er(Dy) with condensate fractions of about 45\,\%. $\tilde{x}$~denotes the horizontal axis perpendicular to the imaging axis. (d--e) Density profiles integrated along $z$, extracted from (c). Solid lines depict the 1D bimodal fit, the dotted lines show gaussian fits to the thermal components. \mbox{(f--h)
    Binary} BECs with a controlled number imbalance giving about $N^\text{Er} = (3.2,6.4,9.2) \times \num{e4}$ with $(35,70,85)\%$ condensate fraction and $N^\text{Dy} = (3.1,2.9,0.9) \times \num{e4}$ with $(30,55,30)\%$ condensate fraction for $(\text{f},\text{g},\text{h})$ respectively. The deformations and the relative displacement of the clouds are caused by inter-species interaction, see main text. The colorbar indicates the optical density.
    }
    \label{fig:degeneracy} 
\end{figure}

We here report on the first experimental realization of quantum-degenerate dipolar mixtures of Er and Dy atoms, using an all-optical approach for trapping and cooling. Er and Dy both have a large magnetic moment of  $7$ and $10$~Bohr magneton, respectively, as well as several stable bosonic and fermionic isotopes with high natural abundances above 15\,\%. Taking advantage of this diversity, we produce  dipolar Bose-Bose mixtures with five different isotope combinations, as well as one Bose-Fermi mixture.  We note that, prior to this work, the production of a nearly doubly degenerate dipolar Bose-Fermi mixture has been reported using two different Dy isotopes \cite{Lu2012} and, more recently, a deeply degenerate Fermi-Fermi mixture has been created from two Er spin states  \cite{Baier2018}. Experimental efforts are also devoted to creating Dy-K mixtures \cite{Grimm2018}.

\begin{figure}[t]
    \includegraphics[]{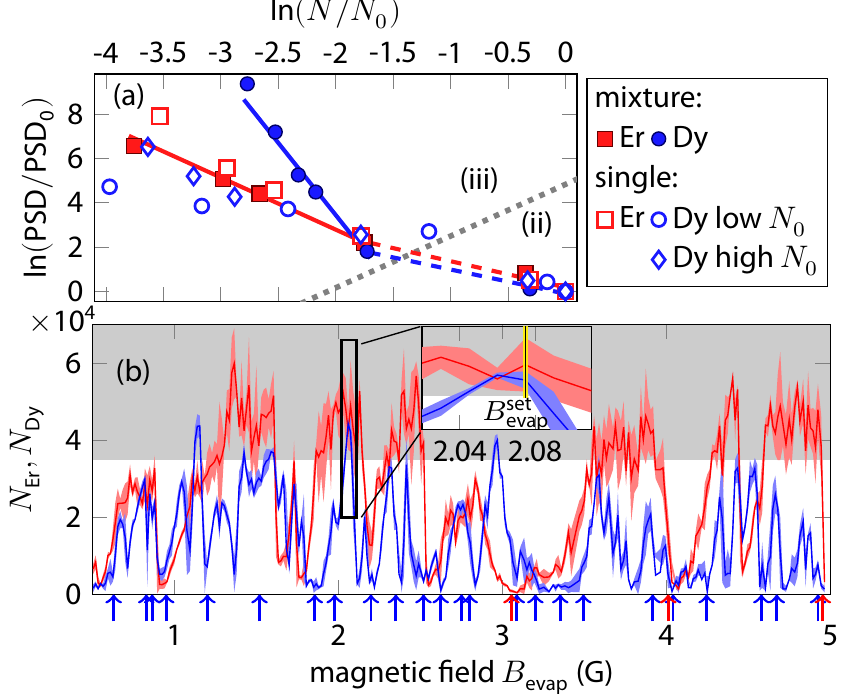}
    \caption{(a) Evaporation trajectories: PSD/PSD\textsubscript{0} as a function of $N/N_0$. Filled squares(circles) indicate the Er(Dy) trajectory in mixture operation. The lines are linear fits to the data for evaporation in the single-beam (ii) and crossed-beam (iii) ODT, see main text. Open symbols show the single-species operation for Er (squares), and for Dy with small (circles) and large (diamonds) initial atom numbers. In the latter case, Dy condenses alone. (b) Atom numbers in the mixture of Er (red) and Dy (blue) at the onset of condensation as a function of the magnetic-field value during evaporation. Condensation is reached for atom numbers above about \num{3.5e4} (gray region). We record the best performance for a~ddBEC around \unit 2.075 G . Arrows indicate the position of known single-species Feshbach resonances \cite{Frisch2014,Baumann2014,Maier2015}.
    }
    \label{fig:evaporation}
\end{figure}

In the following, we detail the production of a double dipolar Bose-Einstein condensate (ddBEC) of \ele{166}{Er} and \ele{164}{Dy}. The same procedure is used for the other isotope mixtures. Our experiment starts with a double  magneto-optical trap (MOT) of Er and Dy, as reported in our recent work~\cite{Ilzhoefer2018}. For both species, the MOT operates on narrow inter-combination lines and yields cold and spin-polarized samples in the absolute lowest Zeeman sublevels  \cite{Frisch2012,Maier2014,Dreon2017,Ilzhoefer2018}. After loading the double MOT, we optically compress the mixture in \unit 400 ms (cMOT phase) by reducing the detuning and power of the MOT beams as well as the magnetic field gradient.

We then transfer the mixture into an optical dipole trap (ODT) by superposing it with the cMOT for \unit 100 ms . Initially, the ODT consists of a  single laser beam at \unit 1064 nm , propagating along the horizontal ($y$) axis. The beam has a fixed vertical ($z$) focus of about~\unit 22 {\mum } , whereas the horizontal waist can be controlled via a time-averaging-potential technique; see e.\,g.~\cite{Baier2012thesis}. This leads to an elliptic beam with variable aspect ratio (AR).  Best transfer efficiency is observed for a beam power  of \unit 32 W and an AR of 4, which provides good spatial overlap between the cMOT and the ODT. We then switch off the MOT beams and magnetic field gradient, and start a 5-seconds evaporation sequence, during which we apply a bias magnetic field $B_\text{evap}$ along the gravity ($z$) axis to preserve spin-polarization. 

Our strategy for evaporative cooling can be divided into three main stages: (i) During the initial \unit 600 ms , we reduce the AR to unity while lowering the power of the single-beam ODT. This increases the density of the mixture at a roughly constant trap depth. (ii) We start forced evaporation in the horizontal ODT and add a vertically-propagating dipole trap beam. The vertical beam is derived from the same laser source as the horizontal one, has a power of \unit 15 W and a waist of \unit 130 {\micro m} . (iii) We proceed with forced evaporation in the crossed ODT by reducing the powers of both beams nearly exponentially until the mixture is close to quantum degeneracy. In the final stage of the evaporation, we increase the AR to 5 to create a pancake-like trapping geometry, and further decrease the trap depth until we reach double quantum degeneracy. To probe the atomic mixture, we switch off the ODT and, after a time-of-flight (TOF) expansion of \unit 25 ms,  we perform sequential absorption imaging with a resonant light pulse at \unit 401 nm for Er and \unit 500 {\micro s } later at \unit 421 nm for Dy~\cite{Aikawa2012,Lu2010}; both pulses have a duration of \unit 50 {\micro s } . The imaging light propagates horizontally with an angle of 45\textdegree~with respect to the $y$-axis. 

Unlike many alkali mixtures \cite{Modugno2002,Hadzibabic2002,Mudrich2002,Grobner2016}, Er and Dy exhibit very comparable atomic polarizabilities, $\alpha$, because of their similar atomic spectra. From single-species experiments \cite{Becher2018,Ravensbergen2018}, a ratio \mbox{$\alpha_\text{Dy}/\alpha_\text{Er} = 1.06$} at \unit 1064 nm is expected. 
For our initial ODT parameters, we calculate trap frequencies of about $\nu^{\text{Er}} = (490,5,1980)\,\si{Hz}$ and $\nu^\text{Dy} = (505,5,2050)\,\si{Hz}$ \footnote{We use the notation $\nu^{\text{Er,Dy}} = (\nu_x,\nu_y,\nu_z)$.}, corresponding to trap depths of \unit 380 {\micro K} and \unit 410 {\micro K } for Er and Dy, respectively. Although small, the difference in trap depths has an important effect on the evaporation trajectory of the mixture. We observe that the more weakly trapped Er atoms act as a coolant for Dy and are preferentially evaporated from the trap (``sympathetic losses'' \cite{Mosk2001,Mudrich2002}). To sustain Er atom numbers high enough to achieve double quantum degeneracy, we imbalance the initial atom number in the MOT with Er as the majority component. 
The atom-number imbalance can be easily controlled by individually changing the MOT loading time and beam power. This strategy is often employed in multi-species experiments,~e.\,g.~\cite{Tang2015,Wu2011}.

\Myref[(a--c)]{fig:degeneracy} shows the phase transition from a thermal Er-Dy mixture (a) to a ddBEC (c).
The TOF absorption images reveal the textbook-like fingerprint of condensation, the emergence of a bimodal density distribution, as plotted in \myref[(d)]{fig:degeneracy}. The condensation series (\myref[(a--c)]{fig:degeneracy}) is taken for an Er(Dy) MOT loading time of \unit 3 s (\unit 1 s), for which we transfer \num{8e6} (\num{7e5}) Er(Dy) atoms into our ODT and measure a temperature of about \unit 35 {\micro K} ; this parameter set allows us to create number-balanced ddBECs. In agreement with the expected polarizabilities, we measure ODT trap frequencies of $\nu^\text{Er} = (50,29.7(9),144(1))\,\text{Hz}$ and $\nu^\text{Dy} = (51,30.2(9),160(1))\,\text{Hz}$~\footnote{The trap frequencies $\nu_y$ and $\nu_z$ are measured from center-of-mass oscillations, whereas $\nu_x$ is extracted from calculations.}. The resultant gravitational sag between the two species is \unit 2.1(2) {\mum} .
By varying the imbalance of the MOT loading, we can produce degenerate mixtures with different atom number ratios and condensate fractions, which is exemplified in \myref[(e,f)]{fig:degeneracy}. For large condensates, one directly observes a deformation of the density profiles due to interspecies interaction, as we discuss later in more detail.

To quantify the cooling efficiency, we plot the normalized phase-space density (PSD/PSD\textsubscript{0}) as a function of normalized atom numbers ($N/N_0$) during the evaporation stages (ii) and (iii), see  \myref[(a)]{fig:evaporation}. PSD\textsubscript{0} and $N_0$ are the respective initial values at stage (ii).
From this plot, we extract $\gamma = -\text{d}\ln(\text{PSD/PSD}_0) / \text{d}\ln(N/N_0)$ \cite{Ketterle1996}, which captures the evaporation efficiency, via a linear fit to the data. 
In the single-beam ODT (stage ii), we see similar efficiencies both in mixture and single-species operation, with $\gamma \approx 1.2$. In the crossed ODT (stage iii), we find $\gamma^\text{Er} = \num{2.4(9)}$ for Er in mixture operation. This value is comparable to state-of-the-art single species Er experiments  \cite{Frisch2014thesis}, and, as expected, little affected by a small admixture of Dy atoms. Contrarily, the cooling efficiency of Dy in stage (iii) strongly benefits from the sympathetic cooling by Er: We observe a steep increase of the Dy PSD in the mixture and extract $\gamma_\text{sym}^{\text{Dy}} = \num{7(2)}$, whereas for the same $N_0^\text{Dy}$ but in single-species operation, the evaporation efficiency is considerably lower and would not suffice for condensation. However, with higher $N_0^\text{Dy}$ we can still produce large Dy BECs in single-species operation.

The proper choice of $B_\text{evap}$ plays an important role for cooling magnetic rare-earth atoms and becomes even more critical in mixture operation. It has indeed been observed in single-species experiments~\cite{Frisch2014,Baumann2014,Maier2015} that both Er and Dy exhibit extremely dense and temperature-dependent spectra of homonuclear Feshbach resonances. \Myref[(b)]{fig:evaporation} shows the atom numbers of the \ele{166}{Er}-\ele{164}{Dy} mixture at the onset of condensation as a function of $B_\text{evap}$ in a small magnetic-field range from  0.5~to~\unit 5 G . As expected, we find a number of broad and narrow loss features. Some of them are connected to known homonuclear Feshbach resonances~\cite{Frisch2014,Baumann2014,Maier2015}, others we attribute to unknown high-temperature resonances or detrimental interspecies scattering conditions. In a few narrow magnetic-field windows we observe atom numbers large enough for both components to condense. However, our magnetic-field stability of about \unit 2 mG is sufficient to reliably operate in most of these small windows. The optimal value of $B_\text{evap}$, listed in \Mytabref{tab:evap_settings}, depends on the isotope combination.

\begin{figure}[t]
\includegraphics[]{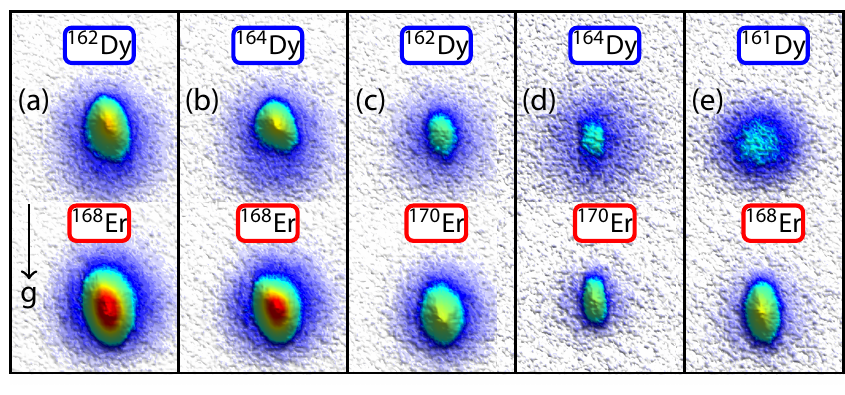}
\caption{Absorption pictures of the double-degenerate Bose-Bose mixtures (a--e) and the Bose-Fermi mixture~(e). The pictures are averaged over 5 to 10 single shots. For all combinations, degeneracy is reached with the evaporation ramp optimized for the \ele{166}{Er}-\ele{164}{Dy} mixture. $B_\text{evap}$ is listed in \Mytabref{tab:evap_settings}. Typical condensate fractions are around 30\,\%, total atom numbers range between \num{1e4} and \num{3.5e4} atoms. For the imbalanced case, higher condensate fractions can be achieved, see \myref{fig:degeneracy}. For the \ele{161}{Dy} Fermi gas, $N=\num{8e3}$, $T/T_F \approx 0.5$, and TOF $=\unit 15 ms $.}
    \label{fig:abspics}
\end{figure}

\begin{table}[b]
\begin{ruledtabular}
\caption{(left) List of optimal $B_\text{evap}$ and $\gamma^\text{Dy}_\text{sym}$ for the quantum degenerate Er-Dy mixtures. (right) Chart of the available isotope mixtures: (\checkmark) realized double-degenerate mixtures, (X)~thermal mixtures, where degeneracy is not yet reached. (\textbf{--})~Mixtures with \ele{167}{Er} and \ele{163}{Dy} are not investigated here.
\label{tab:evap_settings}}
\begin{tabular}{ccrr}
mixture & \!$B_\text{evap}$(G)\! & $\gamma_\text{sym}^\text{Dy}$\vspace*{.5mm}\\
\colrule \\[-2ex]
\ele{166}{Er}-\ele{164}{Dy} & 2.075 &  7(2)& \multirow{6}{*}{\includegraphics[trim = 3 0 0 21.1]{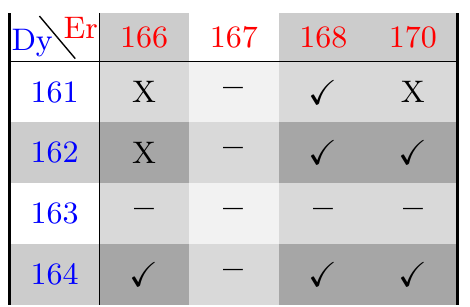}}\\ 
\ele{168}{Er}-\ele{162}{Dy} & 3.300 &  6(2)\\
\ele{168}{Er}-\ele{164}{Dy} & 3.300 &  6(2)\\ 
\ele{170}{Er}-\ele{162}{Dy} & 1.540 &  11(7)\\
\ele{170}{Er}-\ele{164}{Dy} & 3.210 &  3(1)\\
\ele{168}{Er}-\ele{161}{Dy} & 3.455 &  4(1)\\
\end{tabular}
\end{ruledtabular}
\end{table}

Combining Er and Dy offers an unprecedented variety of heteronuclear mixtures with 16 possible isotope configurations, including Bose-Bose, Bose-Fermi, and Fermi-Fermi quantum gases; see \Mytabref{tab:evap_settings}. Using the cooling and trapping procedure optimized for \ele{166}{Er}-\ele{164}{Dy}, we are able to produce five ddBECs and one Bose-Fermi mixture. Concerning the remaining combinations, we know from previous experiments that both \ele{167}{Er} and \ele{163}{Dy} need a different experimental approach since \ele{167}{Er} undergoes light-induced losses in a 1064-nm ODT \cite{Aikawa2014a}, whereas \ele{163}{Dy}, never brought to quantum degeneracy so far, has an inverted hyperfine structure, requiring most probably additional optical pumping stages. Both isotopes will be investigated for future studies of Fermi-Fermi mixtures.

\Myref[(a--d)]{fig:abspics} shows absorption pictures of our doubly degenerate isotope mixtures.  We are able to condense all Bose-Bose isotope mixtures with the exception of \ele{166}{Er}-\ele{162}{Dy}, for which we record severe losses during the evaporation, potentially due to a very large interspecies scattering length. For all degenerate mixtures, we observe sympathetic cooling of Dy by Er. The atom numbers in the ddBECs differ significantly for the different mixtures, while the initial atom numbers in the MOT are very similar. This points to different intra- and interspecies scattering properties during evaporation. The optimal $B_\text{evap}$ and the extracted $\gamma_\text{sym}^{\text{Dy}}$ are listed in~\Mytabref{tab:evap_settings}.

We also prepare one Bose-Fermi mixture, in which a \ele{168}{Er} BEC coexists with a degenerate Fermi gas of \ele{161}{Dy}. 
Although the cooling process of spin-polarized fermions can differ substantially  from bosons, we are able to reach Bose-Fermi degeneracy with a similar evaporation scheme~\footnote{For the Bose-Fermi mixture, slightly different final trap parameters are required, with $\nu^\text{Er} = (33,95(1),120(1)), \nu^\text{Dy} = (34,106(1),142(1))$.}. We measure a temperature of the Fermi gas of $T/T_F \approx 0.5$, with the Fermi temperature $T_F = \unit 140 nK $. We expect that deeper degeneracy might be reached by using smaller ODT beam waists \cite{Aikawa2014a}.

Remarkably, in the TOF images in \myref{fig:degeneracy} and \myref{fig:abspics} hints of interspecies interactions can be spotted: In mixture operation, the center-of-mass (COM) position of each BEC is vertically displaced with respect to its thermal-cloud center; see also \myref[(a)]{fig:oscbyrem}. The two BECs are displaced in opposite direction, with the heavier(lighter) Er(Dy) always shifted down(up). Contrarily, in single-species operation the condensates and thermal clouds are centered; see \myref[(b,c)]{fig:oscbyrem}.
This indicates that the trapped condensates experience a strong repulsive mean-field potential caused by interaction with the other species, whereas the thermal clouds are little affected because of their lower densities. 

To confirm that the displacement after TOF originates from in-trap interspecies interaction, we prepare a ddBEC, let it equilibrate for \unit 50 ms , and then selectively remove either of the two species from the ODT using a resonant light pulse~\footnote{The light pulse at the respective imaging transition has a duration of \unit 1 ms . We have checked that the resonant light of one species does not affect the other.}. After a variable hold time in the ODT, we release the remaining cloud and record its COM position after TOF. As shown in \myref[(d,e)]{fig:oscbyrem}, we observe a very pronounced COM oscillation of the remaining BEC component with a frequency close to its bare trap frequency. The oscillation of Er (removing Dy,~(d)) and of Dy (removing Er, (e)) proceed in counter-phase, as expected from their initial separation in trap. Repeating the same measurement with a thermal-thermal mixture, or a mixture with just one condensed component (not shown), yields negligible or significantly weaker oscillations, respectively.

The relative cloud displacement and the amplitude of the oscillations might in the future serve as a sensitive probe  to extract the overall contact and dipolar interspecies interaction, including the Er-Dy scattering length, which is presently unknown. For this, however, dedicated experiments will have to be combined with simulations based on a generalized coupled Gross-Pitaeskvii equation including interspecies DDI. Such an interaction breaks the angular symmetry of the mean-field interspecies potentials and is expected to render the miscibility-immiscibility boundaries anisotropic and trap-dependent \cite{Wilson2012}.

In conclusion, we have produced heteronuclear dipolar quantum mixtures by combining two strongly-magnetic atomic species, Er and Dy. Their isotope variety, the richness of their interactions, the imbalance in the dipolar strength, and simple laser-cooling schemes make Er-Dy mixtures a powerful experimental platform to access many-body quantum phenomena,  in which contact and dipolar intra- and interspecies interactions are at play.

\begin{figure}[t]
    \includegraphics[]{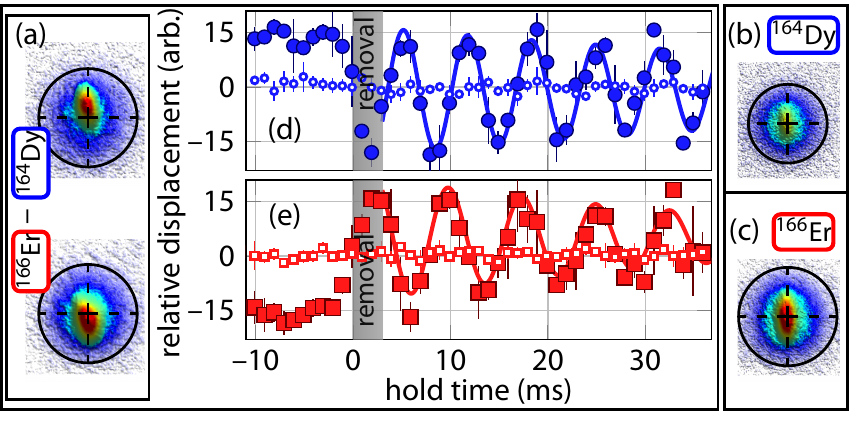}
    
    \caption{Evidence of interspecies interactions in the \ele{166}{Er}-\ele{164}{Dy} mixture: Absorption pictures of Er and Dy in mixture (a) and single-species (b,c) operation. (d,e) Filled symbols show the COM position along $z$ of the Dy BEC (d) and the Er BEC (e) after removal of the other species with resonant light. The gray region indicates the transient time until full removal. The solid lines are damped sine fits to the oscillations. For comparison, open symbols show the COM position in a thermal mixture.}
    \label{fig:oscbyrem}
\end{figure}

This work is supported by the ERC Consolidator Grant (RARE, no.~681432) and a NFRI Grant (MIRARE, no.~\"OAW0600) from the Austrian Academy of Science. G.\,D.~and M.\,S.~acknowledge support by the Austrian Science Fund FWF within the DK-ALM: W1259-N27. We thank the ERBIUM team and the Dy-K team in Innsbruck and the ERBIUM team of Markus Greiner for fruitful discussions.

\end{document}